\documentstyle[epsf,psfig,12pt]{article}
\renewcommand{\baselinestretch}{1.05}

\begin{document}
\def\be{\begin{eqnarray}}
\def\en{\end{eqnarray}}
\def\up{\uparrow}
\def\dw{\downarrow}
\def\non{\nonumber}
\def\la{\langle}
\def\ra{\rangle}
\def\ep{\varepsilon}
\def\vma{{_{V-A}}}
\def\vpa{{_{V+A}}}
\def\nc{N^{\rm eff}_c}
\def\m{\hat{m}}
\def\fp{{f_{\eta'}^{(\bar cc)}}}
\def\lsim{ {\ \lower-1.2pt\vbox{\hbox{\rlap{$<$}\lower5pt\vbox{\hbox{$\sim$}
}}}\ } }
\def\half{{{1\over 2}}}
\def\pr{{\sl Phys. Rev.}~}
\def\prl{{\sl Phys. Rev. Lett.}~}
\def\pl{{\sl Phys. Lett.}~}
\def\np{{\sl Nucl. Phys.}~}
\def\zp{{\sl Z. Phys.}~}

\font\el=cmbx10 scaled \magstep2

\vskip 1.5 cm

\centerline{\bf Recent Progress in Exclusive Charmless Hadronic $B$ Decays:}
\centerline{\bf Status of $B\to\eta' K$ Decay}
\medskip
\bigskip
\medskip
\centerline{\bf Hai-Yang Cheng}
\medskip
\centerline{ Institute of Physics, Academia Sinica}
\centerline{Taipei, Taiwan 115, Republic of China}
\medskip
\bigskip
\bigskip
{\small   
Recent progress in exclusive charmless hadronic decays of the $B$ meson
is discussed.}

\noindent {\bf 1.~Introduction}

Recently there has been a remarkable progress in the study of exclusive
charmless $B$ decays, both experimentally and theoretically. On the
experimental side, many new two-body decay modes were discovered by CLEO
\cite{CLEO}:
\be
 B\to\eta' K^+,~\eta' K^0_S,~\pi^+ K^0_S,~\pi^0 K^+,~\omega K^+,~\omega 
K^0_S,~\omega\pi^+,~\phi K^*.
\en
Moreover, CLEO has improved upper limits for many other channels. Therefore,
it is a field whose time has finally arrived. On the theoretical aspect, 
there are two important issues to be addressed: (i) the renormalization 
scheme and scale dependence of hadronic matrix elements, and (ii) 
nonfactorizable effects in charmless $B$ decays. A fascinating
progress in dealing with the above-mentioned theoretical issues has been made
over the last few years. In this talk
I'll first discuss the theoretical progress and then proceed to elaborate
the decay $B\to\eta' K$ which has received a lot of attention recently.

\noindent {\bf 2.~Renormalization scale and scheme dependence of hadronic 
matrix element}

The relevant effective weak Hamiltonian for hadronic weak $B$ decay is of
the form
\be
{\cal H}_{\rm eff}={G_F\over\sqrt{2}}\left[ \lambda_u(c_1O_1^u+c_2O_2^u)
+\lambda_c(c_1O_1^c+c_2O_2^c)-\lambda_t\sum_{i=3}^{10} c_iO_i\right],
\en
where $\lambda_u=V_{ub}V^*_{uq}$.
The Wilson coefficients $c_i(\mu)$ in Eq.~(2) have been evaluated at the
renormalization scale $\mu\sim m_b$ to the next-to-leading order. Beyond the
leading logarithmic approximation, they depend on the choice of the
renormalization scheme. The mesonic matrix
elements are customarily evaluated under the factorization hypothesis. In the
naive factorization approach, the relevant Wilson coefficient functions
for color-allowed external $W$-emission (or so-called ``class-I") and 
color-suppressed (class-II) internal $W$-emission
amplitudes are given by $a_1=c_1+c_2/N_c$, $a_2=c_2+c_1/N_c$, respectively, 
with $N_c$ the number of colors. Inspite of its tremendous simplicity, naive
factorization encounters two major difficulties. First, it never works for
the decay rate of class-II decay modes, though it usually operates for
class-I transition. 
Second, the hadronic matrix element under factorization is
renormalization scale $\mu$ independent as the vector or axial-vector
current is partially conserved. Consequently, the amplitude $c_i(\mu)
\la O\ra_{\rm fact}$ is not truly physical as the scale dependence of
Wilson coefficients does not get compensation from the matrix elements.
The first difficulty indicates that it is inevitable and 
mandatory to take into account nonfactorizable contributions, especially 
for class-II decays, to render the color suppression of internal $W$
emission ineffective. The second difficulty also should
not occur since the matrix elements of four-quark 
operators ought to be evaluated in the same renormalization scheme as that for 
Wilson coefficients and renormalized at the same scale $\mu$. 

   To circumvent the aforementioned second problem, one should evaluate 
perturbative QCD and electroweak corrections to the hadronic weak matrix
elements parametrized by the matrices $\m_s$ and $\m_e$, respectively,
so that \cite{Buras92}
\be
c_i(\mu)\la O_i(\mu)\ra=\,c_i(\mu)\left[\,{\rm I}+{\alpha_s(\mu)\over 4\pi}
\m_s(\mu)+{\alpha\over 4\pi}\m_e(\mu)\right]_{ij}\la O_j^{\rm tree}\ra\equiv
c_j^{\rm eff}\la O_j^{\rm tree}\ra.
\en
Then factorization is applied to the matrix elements of tree operators;
that is, before employing factorization to the four-quark operators, it is
necessary to absorb all corrections to $O^{\rm tree}$ into the effective
coefficients $c_i^{\rm eff}$. 
One-loop penguin corrections and vertex corrections to the operators $O_i$ 
have been calculated in \cite{Buras92,Kramer,Ali}. One can explicitly
check that the effective Wilson coefficients $c_i^{\rm eff}$ are indeed
renormalization scheme independent and approximately renormalization scale
independent \cite{CT97a}. It should be stressed that the effective penguin 
coefficients $c_i^{\rm eff}$ ($i=3,\cdots,10$) take into account 
the effect of penguin diagrams not only with top quark exchange 
characterized by the Wilson coefficients $c_i$ but also with internal $u$ 
and $c$ quarks induced by the tree operators $O_1^u$ and $O_1^c$, respectively.

\noindent {\bf 3.~Nonfactorizable effects in charmless hadronic $B$ decay}

 As stressed in the last section, it is mandatory to take into account the
nonfactorizable effects, especially for class-II modes.
For $B\to PP$ or $VP$ decays, 
nonfactorizable contributions can be lumped into the effective parameters
$a_1$ and $a_2$ \cite{Cheng94}:
\be
a_1^{\rm eff}=c_1+c_2\left({1\over N_c}+\chi_1\right), \qquad
a_2^{\rm eff}=c_2+c_1\left({1\over N_c}+\chi_2\right), 
\en
where $\chi_i$ are nonfactorizable terms and receive main contributions from
the color-octet current operators. Phenomenological analyses
of two-body decay data of $D$ and $B$ mesons indicate that while
the generalized factorization hypothesis in general works reasonably well, 
the effective parameters $a_{1,2}^{\rm eff}$ do show some variation from
channel to channel, especially for the weak decays of charmed mesons
 \cite{Cheng94,Kamal96}.
An eminent feature emerged from the data analysis is that $a_2^{\rm eff}$ 
is negative in charm decay, whereas it becomes positive in bottom decay 
\cite{Cheng94,CT95,Neubert}:
\be
a_2^{\rm eff}(D\to\bar K\pi)\sim -0.50\,, \quad a_2^{\rm eff}(B\to D\pi)\sim 
0.26\,,
\en
which in turn implies
\be
\chi_2(\mu\sim m_c;~D\to \bar K\pi)\sim -0.36\,, \quad \chi_2(\mu\sim m_b;~B
\to D\pi)\sim 0.11\,.
\en
The observation $|\chi_2(B)|\ll|\chi_2(D)|$ is consistent
with the intuitive picture that soft gluon effects become stronger when the
final-state particles move slower, allowing more time for significant
final-state interactions after hadronization \cite{Cheng94}.
Phenomenologically, it is often to treat the number of colors $N_c$ as
a free parameter and fit it to the data. Theoretically, this amounts to
defining an effective number of colors by $1/N_c^{\rm eff}\equiv 
(1/N_c)+\chi$. It is clear from Eq.~(6) that
\be
N_c^{\rm eff}(D\to \bar K\pi) \gg 3,\quad N_c^{\rm eff}(B\to D\pi)
\sim 2.
\en

   It is natural to ask that does the naive factorization approach also fail
in charmless $B$ decays ? If so, how large is the nonfactorizable effect ?
Since the energy release in charmless two-body decays of the $B$ meson 
is generally 
slightly larger than that in $B\to D^{(*)}\pi$, it is expected
that $\nc$ for the $B$ decay into two light mesons is close
to $\nc(B\to D\pi)\approx 2$. 
It is pointed out in \cite{Deandrea} that the parameters
$a_2,~a_3$ and $a_5$ are strongly dependent on $\nc$ and the rates dominated
by these coefficients can have large variation. For example, the decay widths
of $B^-\to \omega K^{(*)-}$, $B^0\to\omega K^0,~\rho K^{*0}$, $B_s\to\eta
\omega,~\eta\phi,~\omega\phi,\cdots,$ etc. have strong $N_c$ dependence 
\cite{Deandrea}. We have shown recently in \cite{CT97b} that the branching 
ratio of $B^-\to\omega K^-$ has its lowest value of order $1\times 10^{-6}$
near $\nc\sim 3-4$ and hence the 
naive factorization with $\nc=3$ is ruled out by experiment,
${\cal B}(B^\pm\to\omega K^\pm)=\left(1.5^{+0.7}_{-0.6}\pm 0.3\right)
\times 10^{-5}$ \cite{CLEO}.

However, it is not easy to discern 
between $\nc=\infty$ and $\nc=2$ in $B\to\omega K$ decays, and it becomes
important to have a 
more decisive test on $\nc$. For this purpose, we shall focus 
on the decay modes dominated by the tree diagrams and sensitive to
the interference between external and internal $W$-emission amplitudes.
The fact that $\nc=2$ ($\nc=\infty$) implies constructive (destructive) 
interference will enable us to differentiate between them. Good
examples are the class-III modes: $B^\pm\to \pi^0\pi^\pm,~\eta\pi^\pm,~\pi^0
\rho^\pm,~\omega\pi^\pm,\cdots$. We found that \cite{CT97b}  
the averaged branching ratio of $B^\pm\to\omega\pi^\pm$ has its
lowest value of order $2\times 10^{-6}$ at $\nc=\infty$ and then increases 
with $1/\nc$. Since experimentally \cite{CLEO}
${\cal B}(B^\pm\to\omega\pi^\pm)=\left(1.1^{+0.6}_{-0.5}\pm 0.2\right)
\times 10^{-5}$, 
it is evident that $\nc=\infty$ is disfavored by the data.
Measurements of interference effects in charged $B$ decays
$B^-\to\pi^-(\rho^-)\pi^0(\rho^0)$ will help determine the parameter $\nc$.
For example, the ratio $R_1\equiv 2{\cal B}(B^-\to\pi^-\rho^0)/{\cal B}
(\bar B^0\to\pi^-\rho^+)$ is calculated to be 2.50 for $\nc=2$ and 0.26
for $\nc=\infty$. Hence, a measurement of $R_1$, which has the advantage of
being independent of the Wolfenstein parameters $\rho$ and $\eta$,
will provide a decisive determination of the effective number of colors $\nc$.

\noindent {\bf 4.~Exclusive charmless hadronic $B$ decays to $\eta'$ and 
$\eta$}

The CLEO collaboration has recently reported the preliminary branching ratios 
for the exclusive decay $B\to\eta'K$ dominated by gluonic penguin
daigrams \cite{CLEO}:
\be
&& {\cal B}(B^\pm\to\eta'K^\pm) \equiv {1\over 2}\left[{\cal B}(B^+\to\eta' 
K^+)+{\cal B}(B^-\to\eta' K^-)\right]
= \left(7.1^{+2.5}_{-2.1}\pm 0.9\right)
\times 10^{-5}, \non\\
&& {\cal B}(\stackrel{_{_{(-)}}}{B}\!\!{^0}\to\eta' \stackrel{_{_{(-)}}}{K}
\!\!{^0}) \equiv {1\over 2}\left[{\cal B}(B^0\to\eta' K^0)
+{\cal B}(\bar B^0\to\eta' \bar K^0)\right]
= \left(5.3^{+2.8}_{-2.2}\pm 1.2\right)\times 
10^{-5}.
\en
Early theoretical estimate of the $B^\pm\to\eta' K^\pm$ branching ratio 
\cite{Chau1,Du,Kramer} lies in the range of $(1-2)\times 10^{-5}$. 
\footnote{The prediction ${\cal B}(B^\pm\to\eta' K^\pm)=3.6\times 10^{-5}$
given in \cite{Chau1}
is too large by about a factor of 2 because the normalization constant (i.e.
$1/\sqrt{3}$) of the $\eta_0$ wave function was not taken into account in the 
form factor $F_0^{B\eta_0}$. This negligence was also made in some recent 
papers on $B\to\eta' K$.}
The CLEO result thus appears to be abnormally large. The question is then 
can the CLEO observation of $B\to \eta' K$ be accommodated in the standard 
model ? Do the new data imply new physics ? The theoretical interest 
and speculation in this
subject has surged, as evidenced by the recent literature [4,14-20] that 
offer various interpretations on the unexpected large branching ratios.

In order to illustrate the problem clearly we choose the
following parameters for calculation:
\be
F_0^{BK}(0)=0.34,  \quad
\sqrt{3}F_0^{B\eta_0}(0)=0.254, \quad f_0=f_8=f_\pi, 
\quad m_s=150\,{\rm MeV}, \quad \theta=-19.5^\circ,  
\en
and $\nc=2$, where $\theta$ is the $\eta-\eta'$ mixing angle.
Using the renormailization scale and scheme independent effective Wilson
coefficients $c_i^{\rm eff}$ discussed in Sec.~II, we find
\be
{\cal B}(B^\pm\to \eta' K^\pm)=\cases{ 1.4\times 10^{-5},& for~$\eta=0.35$,
~$\rho=0.08$,  \cr 1.6\times 10^{-5},& for~$\eta=0.34$, ~$\rho=-0.12$. \cr}
\en
In the ensuing discussion, we will use (9) and (10) as the benchmarked 
values to be compared with. 
  Since the choice of form factors and light quark masses is uncertain, one 
may argue that the CLEO data (8) can be fitted by choosing
a small strange quark mass and/or large form factors $F_0^{BK}$ 
and $F_0^{B\eta_0}$ \cite{Datta}. For example,
the above model estimate with $m_s=55$ MeV or $F_0^{BK}(0)=0.63$ will 
fit to the central value of ${\cal B}(B^\pm\to\eta' K^\pm)$. However it is 
dangerous to fit the parameters to a few particular
decay modes. The point is that comparison between theory and experiment
should be carried out using the same set of parameters for all decay
channels. Indeed the measured branching ratio of $B\to \pi K$ puts a 
constraint on the strange quark mass and it indicates that $m_s$ cannot be too 
small. In the SU(3) limit we have the relation $F_0^{B\pi^\pm}=F_0^{BK}$. 
Most of the existing QCD-sum-rule and quark model calculations show that 
$F_0^{B\pi^\pm}(0)\lsim 0.33$ (for a review, see \cite{Cas}). We shall see
below that a severe constraint on $F_0^{B\pi^\pm}
(0)$ can be derived from the current limit on the decay $B^+\to\eta\pi^+$.
Since SU(3) breaking is expected at most of 30\% level,
it is very unlikely that $F_0^{BK}(0)$ can deviate much from 0.33\,. 
Likewise, the nonet symmetry relation $\sqrt{3}F_0^{B\eta_0}=F_0^{B\pi^\pm}$
implies that $\sqrt{3}F_0^{B\eta_0}(0)$ cannot be too large than the model
estimate, say 0.254 given in (9).
In short, the parameters given in (9) cannot be modified
dramatically without violating SU(3) symmetry relation and experimental
observation of other decay modes.

   Nevertheless, we can adjust the parameters in (9) slightly to improve the 
discrepancy between theory and experiment. The key point is that an
accummulation of several small enhancement may eventually lead to a sizable 
enhancement. First, the current quark mass $m_s=150$ MeV in (9) is
defined at the renormalization scale $\mu=1$ GeV. For reason of consistency,
one ought to apply the small running quark mass $m_s(m_b)\simeq 105$ MeV in 
calculation. Second, we use $f_8/f_\pi=1.38\pm 0.22$ and $f_0/f_\pi=1.06\pm 
0.03$ to take into account SU(3) breaking effects in decay constants 
\cite{Holstein}. Third, for the form factor $F_0^{B\eta_0}$ we follow 
\cite{Ali} to use $\sqrt{3}F_0^{B\eta_0}(0)=0.33$, which is slightly larger 
than the value of 0.254 obtained in \cite{BSW}. Fourth, previously
we employed the mixing angle $\theta=-19.5^\circ$ so that the wave functions
of the $\eta$ and $\eta'$ have the simple expressions: $\eta=
{1\over\sqrt{3}}(u\bar u+d\bar d-s\bar s)$ and $\eta'=
{1\over\sqrt{6}}(u\bar u+d\bar d+2s\bar s)$.
Here we instead use the value $\theta=-22.0^\circ\pm 3.3^\circ$ extracted in 
\cite{Holstein}. Applying the new value for each of the aforementioned 
parameters individually, we 
find that the branching ratio of $B^\pm\to\eta' K^\pm$ is enhanced by 62\%, 
37\%, 19\%, and 5\%, respectively. Hence, the dominant enhancement comes
from the running strange quark mass and SU(3) breaking in decay constants. 
When all new parameters are employed, we obtain (see Table I)
\be
{\cal B}(B^\pm\to \eta' K^\pm)=\cases{ 3.87\times 10^{-5}, \cr 4.04\times 
10^{-5},  \cr} \qquad {\cal B}(\stackrel{_{_{(-)}}}{B}\!\!{^0}\to\eta' 
\stackrel{_{_{(-)}}}{K}\!\!{^0})=
\cases{ 3.70\times 10^{-5}, \cr 3.67\times 10^{-5}, \cr}
\en
for $\eta=0.35$, $\rho=0.08$ (upper entry) and $\eta=0.34$, $\rho=-0.12$
(lower entry), respectively.
Note that ${\cal B}(B^\pm\to\eta' K^\pm)$ increases with $1/\nc$ and it
becomes as large as order of $6\times 10^{-5}$ at $\nc=\infty$. However, as 
we have stressed before, it is very unlikely to have a large $\nc$  
in the two-body charmless $B$ decay. Therefore, we see that by adjusting 
the parameters in (9) within some reasonable range, the standard penguin 
contribution can account for the observed decay rate of $B^0\to\eta' K^0$ but 
only marginally for $B^\pm\to\eta' K^\pm$. Nevertheless, the current data
allow for some new contributions 
(but not necessarily new physics) unique to the $\eta'$. Of course, we have 
to await more new data to sort it out.

   There are several mechanisms which are unique to the $\eta'$ and may 
enhance the decay rate of $B\to\eta' K$. (i) The $b\to sg^*$ penguin 
followed by the transition $g^*\to g\eta'$ via the QCD anomaly can in 
principle contribute to the 
exclusive decay $B\to\eta' K$. This anomalous mechanism was originally
advocated to explain the observed large inclusive $B\to\eta'+X$ signal
\cite{Soni,Hou}.
However, since this mechanism involves a production of a gluon before 
hadronization, it will not play an essential role in low-multiplicity
two-body exclusive decays unless the gluon is soft
and absorbed in the wave function of the $\eta'$. Another possibility is that
the gluon produced from the penguin diagram and the gluon emitted from the
light antiquark fuse into the $\eta'$ \cite{Ahmady,Du97}.
As the average momentum of the gluon emitted from the antiquark is 
in general less than 1 GeV, it is not clear if
perturbative QCD is still applicable in this case. (ii) The process
$b\to s+g^*g^*\to s+\eta'$ involves two gluon production in the penguin-like
diagram followed by the $\eta'$-gluon anomalous interaction. The decay
$b\to s+g^*g^*$ has been calculated in the literature \cite{Wyler}.
It appears that the branching ratio arising from this mechanism is less 
than $1\times 10^{-5}$ \cite{Lin}.
(iii) A new internal $W$-emission contribution comes from 
the Cabibbo-allowed process $b\to c\bar c s$ followed 
by a conversion of the $c\bar c$ pair into the $\eta'$ via two gluon 
exchanges. This new contribution
is important since its mixing angle $V_{cb}V_{cs}^*$ is as large as that
of the penguin amplitude and yet its Wilson coefficient $a_2^{\rm eff}$ 
is larger than that of penguin operators. 
The decay constant $f_{\eta'}^{(\bar cc)}$, 
defined by $\la 0|\bar c\gamma_\mu\gamma_5c|\eta'\ra=if_{\eta'}^{(\bar cc)}
q_\mu$, has been estimated to be $|f_{\eta'}^{(\bar cc)}|=(50-180)$ MeV,
based on the OPE, large-$N_c$ approach and QCD low energy 
theorems \cite{Halp1}. It was claimed in \cite{Halp1,Halp2} 
that $|f_{\eta'}^{(\bar cc)}|\sim 140$ MeV is needed in order to exhaust the 
CLEO observation of $B^\pm\to \eta' K^\pm$ and $B\to\eta'+X$ by the mechanism
$b\to c\bar c+s\to\eta'+s$ via gluon exchanges. However, a large value of 
$\fp$ seems to be ruled out for three reasons. First, the decay constant
$f_{\eta'}^{(\bar uu)}$ is only of order 50 MeV. Second, suppose the 
pseudoscalar content of $c\bar c$ is dominated by the $\eta_c$. Then from the
data of $J/\psi\to\eta_c\gamma$ and $J/\psi\to\eta' \gamma$, one can show
that $|\fp|\geq 6$ MeV, where the lower bound corresponds to the 
nonrelativitsic quark estimate. (When the relativistic effect of the $\eta'$ 
in $J/\psi\to
\eta'\gamma$ is taken into account, $|\fp|$ is larger than 6 MeV.) Even when
contributions from e.g., $\eta'_c$, $\eta_c'',\cdots$ to the $c\bar c$ are
included, it is argued in \cite{Chao} that $|\fp|\leq 40$ MeV. Third, based 
on the $\eta\gamma$ and $\eta'\gamma$ transition form
factor data, the range of allowed $\fp$ is estimated to be $-65\,{\rm MeV}
\leq\fp\leq 15$ MeV \cite{Kroll}.

\vskip 0.4cm
\begin{table}
{{\small Table I. Averaged branching ratios for charmless $B$ decays to 
$\eta'$ and $\eta$, where ``Tree" refers to branching ratios 
from 
tree diagrams only, ``Tree+QCD" from tree and QCD penguin diagrams, and 
``Full" denotes full contributions from tree, QCD and electroweak (EW) penguin
diagrams in conjunction with contributions from the process $c\bar 
c\to\eta_0$. Predictions are for $k^2=m_b^2/2$, $\nc=2$, $\fp=-15$ MeV, 
$\eta=0.35,~\rho=0.08$ (the first number in parentheses) and $\eta=0.34,~\rho
=-0.12$ (the second number in parentheses).}
{\footnotesize
\begin{center}
\begin{tabular}{|l|c c c c |c|} \hline
Decay & Tree & Tree$+$QCD & Tree$+$QCD$+$EW & Full & Exp. \cite{CLEO}\\ 
\hline 
$B^\pm\to\eta' K^\pm$ & $1.41\times 10^{-7}$ & $(4.00,~4.18)\,10^{-5}$ & 
$(3.87,~4.04)\,10^{-5}$ & $(5.48,~5.69)\,10^{-5}$ &
$(7.1^{+2.5}_{-2.1}\pm 0.9)\,10^{-5}$ \\
$B^\pm\to\eta K^\pm$ & $3.56\times 10^{-7}$ & $(5.48,~3.37)\,10^{-7}$ &
$(3.40,~3.80)\,10^{-7}$ & $(5.05,\,8.68)\,10^{-7}$  & $<0.8\times 10^{-5}$ \\
$B^\pm\to\eta' K^{*\pm}$ & $2.11\times 10^{-7}$ & $(2.69,~1.90)\,10^{-6}$ &
$(2.94,~2.11)\,10^{-6}$ & $(5.90,\,3.24)\,10^{-7}$ & $<29\times 10^{-5}$ \\
$B^\pm\to\eta K^{*\pm}$ & $5.25\times 10^{-7}$ & $(0.92,~1.53)\,10^{-6}$ &
$(1.52,~2.42)\,10^{-6}$ & $(2.49,\,3.70)\,10^{-6}$ & $<24\times 10^{-5}$ \\
$B^\pm\to\eta'\pi^{\pm}$ & $1.94\times 10^{-6}$ & $(1.13,~1.06)\,10^{-5}$ &
$(1.12,~1.05)\,10^{-5}$ & $(1.29,\,1.21)\, 10^{-5}$ & $<4.5\times 10^{-5}$ \\
$B^\pm\to\eta \pi^{\pm}$ & $4.93\times 10^{-6}$ & $(9.57,~6.02)\,10^{-6}$ &
$(9.82,~6.24)\,10^{-6}$ & $(1.04,\,0.67)\, 10^{-5}$ & $<0.8\times 10^{-5}$ \\
$B^\pm\to\eta' \rho^{\pm}$ & $3.95\times 10^{-6}$ & $(1.08,~1.84)\,10^{-5}$ &
$(1.08,~1.84)\,10^{-5}$ & $(1.01,\,1.71)\, 10^{-5}$ & \\
$B^\pm\to\eta \rho^{\pm}$ & $9.72\times 10^{-6}$ & $(1.21,~1.71)\,10^{-5}$ &
$(1.19,~1.66)\,10^{-5}$ & $(1.19,\,1.63)\, 10^{-5}$ &  \\
\hline
$ B_d\to\eta' K^0$ & $5.09\times 10^{-9}$ &$(3.82,\,3.80)\, 10^{-5}$
&$(3.70,\,3.67)\, 10^{-5}$ &$(5.22,\,5.19)\, 10^{-5}$ & $(5.3^{+2.8}_{-2.2}
\pm 1.2)\, 10^{-5}$ \\
$ B_d\to\eta K^0$  & $1.93\times 10^{-8}$ & $(1.23,\,0.77)\, 10^{-7}$ 
& $(1.97,\,3.11)\, 10^{-8}$ & $(2.49,\,3.28)\, 10^{-7}$ & \\
$ B_d\to\eta' K^{*0}$ & $3.88\times 10^{-9}$ & $(2.23,\,2.08)\,
10^{-6}$ & $(2.47,\,2.32)\, 10^{-6}$ & $(2.44,\,2.96)\, 10^{-7}$ &
$<4.2\times10^{-5}$ \\ 
$ B_d\to\eta K^{*0}$ & $1.62\times 10^{-8}$ & $(5.71,\,6.64)\,
10^{-7}$  & $(1.23,\,1.36)\, 10^{-6}$  & $(2.26,\,2.44)\, 10^{-6}$  &
$<3.3\times 10^{-5}$ \\ 
$ B_d\to\eta'\pi^{0}$ & $3.31\times 10^{-11}$ & $(4.83,\,6.80)\,
10^{-6}$  & $(4.63,\,6.52)\, 10^{-6}$  & $(5.32,\,7.35)\, 10^{-6}$  & 
$<2.2\times 10^{-5}$ \\
$ B_d\to\eta \pi^{0}$ & $6.70\times 10^{-9}$ & $(2.71,\,3.67)\,
10^{-6}$ & $(2.66,\,3.60)\, 10^{-6}$ & $(2.86,\,3.83)\, 10^{-6}$ & \\
$ B_d\to\eta'\rho^{0}$ & $1.21\times 10^{-7}$ &$(2.07,\,3.33)\,
10^{-6}$ &$(2.01,\,3.24)\, 10^{-6}$ &$(1.79,\,2.93)\, 10^{-6}$ & \\
$ B_d\to\eta \rho^{0}$ & $3.46\times 10^{-7}$ &$(1.28,\,2.20)\,
10^{-6}$ &$(1.16,\,1.98)\, 10^{-6}$ &$(1.11,\,1.89)\, 10^{-6}$ &
$<8.4\times 10^{-5}$   \\
\hline
\end{tabular}
\end{center} } }
\end{table} 
\vskip 0.4cm

  From Table I it is clear that for $\fp=-15$ MeV, which is consistent with
above-mentioned constraints, the agreement between theory and experiment
for $B\to\eta' K$ is substantially improved in the presence of large charm
content in the $\eta'$. We conclude that no new physics is needed to account
for the CLEO data of $B\to\eta' K$.
We have also calculated the branching ratios of other exclusive charmless 
$B$ decays involving $\eta'$ and $\eta$ (see Table I), where 
use of $f_\eta^{(c\bar c)}=-\tan\theta\fp$ has been made. 
\footnote{In the two mixing angle parametrization scheme given in
\cite{Kroll}, the decay constant $f_\eta^{(c\bar c)}$ is much smaller: 
$f_\eta^{(c\bar c)}=-\tan\theta_1\fp$ with $\theta_1=-(6^\circ\sim 9^\circ)$.}
Three comments are in order. (i) The effect of $c\bar c$ conversion into
the $\eta'$ contributes destructively to $B\to\eta' K^*$. Consequently,
the branching ratio of $B\to\eta' K^*$ is suppressed
\footnote{In \cite{CT97a} we have employed $\fp\sim -50$ MeV. In that case, 
$B\to\eta' K^*$ is dominated by the process $c\bar c\to\eta_0$ and its
branching ratio is of order $10^{-5}$.}
and ${\cal B}(B\to\eta' K^*)/{\cal B}(B\to\eta K^*)\sim {\cal O}(10^{-1})$. 
If $B\to\eta' K$ is assumed to be entirely accommodated by any of 
aforementioned new mechamisms, the decay rate of $B\to\eta' K^*$ will 
be predicted
to be the same order of magnitude as $B\to\eta' K$ \cite{Halp2,Ahmady}.
(ii) Contrary to $B\to\eta(\eta')K^*$ decays, we see from Table I that
${\cal B}(B\to\eta' K)/{\cal B}(B\to\eta K)\sim {\cal O}(10^{2})$ due to
the destructive interference in the penguin diagrams of $B\to\eta K$.
(iii) The electroweak penguin effects are in general very small, but they
become important for $B\to\eta K$ and $B\to\eta K^*$ decays due to a large
cancellation of QCD penguin contributions in these decay modes.

For $B\to\eta'(\eta)\pi(\rho)$ decays, the mechanism of $c\bar c\to\eta_0$
is much less dramatic since it does not gain mixing-angle enhancement as in 
the case of $B\to\eta'(\eta)K(K^*)$. Their branching ratios are sensitive
to the light quark masses $m_u$, $m_d$ and form factors such as $F_0^{B\pi}$
\cite{CT97a}.
The current experimental limit on the decay $B^\pm\to\eta\pi^\pm$ puts 
useful constraints 
on $m_q$ and $F_0^{B\pi}$. The predicted values presented in Table I are
for $m_u(m_b)=5$ MeV, $m_d(m_b)=10$ MeV and $F_0^{B\pi^\pm}(0)=0.30$.
We find that even a slight increase of $F_0^{B\pi}(0)$ or decrease of $m_q$,
say $m_u(m_b)\approx m_d(m_b)/2\sim 3$ MeV,  
will make the decay rate of $B^\pm\to\eta\pi^\pm$ exceeding the present 
upper bound significantly. We also see that a negative $\rho$,
which in turn implies a unitarity triangle $\gamma$ in the range 
$90^\circ<\gamma<180^\circ$, is preferred \cite{CT97b}. By contrast, the
present experimental value of the ratio $R_2\equiv\Gamma(B^0\to\pi^\mp K^\pm)/
\Gamma(B^\pm\to\pi^\pm K^0)$ favors a positive $\rho$ \cite{Ali}. Note that
a positive $\rho$ is also preferred by the limit on the ratio $\Delta M_s/
\Delta M_d$ \cite{Pag}. Clearly more data of $B^\pm\to\eta\pi^\pm$ and $R_2$ 
are needed to pin down the sign of $\rho$.

\vskip 0.2cm
\noindent {\bf Acknowledgments}

\noindent I would like to thank the organizers for this well run, fruitful 
and stimulating workshop. 
This work was supported in part by the National 
Science Council of ROC.

%%%%% References %%%%%%%%%%%%%%%%%%%%%%%%%%%%%%%%%%%%%%%%%%%%%%%%%%%%%%%%%%%%%
\renewcommand{\baselinestretch}{1.0}
\newcommand{\bi}{\bibitem}
{\small
%
%\newpage
 }

\newpage

\end{document}